\documentclass[aps,prd,twocolumn,groupedaddress,showpacs]{revtex4-1}

\usepackage{graphicx}
\def \AD {A_D}
\def \ADbar {A_{\bar D}}
\def \AbarD {{\bar A}_D}
\def \AbarDbar {{\bar A}_{\bar D}}

\def \Af {A_f}
\def \Afbar {A_{\bar f}}
\def \Abarf {{\bar A}_f}
\def \Abarfbar {{\bar A}_{\bar f}}

\def \qp {\frac{q}{p}}
\def \pq {\frac{p}{q}}

\def \bmampint {\int^{+\infty}_0|A(B^-\to[f]_DK^-)|^2 dt}

\def \BmAmpOne {\AD\Af+\ADbar\Abarf}
\def \BmAmpTwo {\qp\AD\Abarf+\pq\ADbar\Af}

\def \IntGp {\int^{+\infty}_0|g_+(t)|^2\epsilon(t)\,dt}
\def \IntGm {\int^{+\infty}_0|g_-(t)|^2\epsilon(t)\,dt}
\def \IntReGpGm {\int^{+\infty}_0\Re[g_+(t)g^*_-(t)]\epsilon(t)\,dt}
\def \IntImGpGm {\int^{+\infty}_0\Im[g_+(t)g^*_-(t)]\epsilon(t)\,dt}

\def \rf {r_f e^{-i\delta_f}}

\def \rb {r_B e^{i(\delta_B-\gamma)}}

\def \rbarfbar {\bar r_{\bar f} e^{-i\bar\delta_{\bar f}}}
\def \fcp {f_{CP}}

\def \xb {x_{B\pm}}
\def \yb {y_{B\pm}}
\def \xbprime {x'_{B\pm}}
\def \ybprime {y'_{B\pm}}

\def \btodstkst {B^-\to D^{(*)0}K^{(*)-}}

\def \xyOrb {\sqrt{x^2+y^2}/r_B}

\renewcommand{\Re}{{\rm Re}}
\renewcommand{\Im}{{\rm Im}}

\begin{document}


\title{Effect of $D-\bar D$ mixing in the extraction of $\gamma$ with $B^-\to D^0 K^-$ and $B^-\to D^0\pi^-$ decays}


\author{Matteo Rama}
\affiliation{INFN, Laboratori Nazionali di Frascati, I-00044 Frascati, Italy}


\date{\today}

\begin{abstract}
We examine the impact of $D-\bar D$ mixing in the extraction of the angle $\gamma$ of the unitarity triangle with $\btodstkst$ and $B^-\to D^{(*)0}\pi^-$ decays. We point out that the leading corrections, linear in the small mixing parameters $x=\Delta m/\Gamma$ and $y=\Delta\Gamma/2\Gamma$, depend on how the signal selection efficiency varies as a function of the $D$ proper time, and we estimate the scale factor in a simplified case. We note that the charm mixing effect is suppressed in the GLW method. We discuss the case where the leading corrections become quadratic in $x$ and $y$, and we point out some limitations of this scenario. We compute the bias $\Delta\gamma$ when $D-\bar D$ mixing is ignored in the rates of $\btodstkst$ but not in the measurement of the $D$ decay parameters. We find $|\Delta\gamma|\lesssim 1^\circ$ for all $\btodstkst$ decays, limited by the measured value of the strong phase difference between the $b\to u\bar c s$ and $b\to c\bar u s$ decay amplitudes. On the other hand, we remark that the effect in $B^-\to D^{(*)0}\pi^-$ decays cannot be neglected.
\end{abstract}


\maketitle

\section{Introduction}
The most sensitive method used so far to measure the angle $\gamma$ ($\phi_3$) of the unitarity triangle exploits the interference between the $b\to c\bar u s$ and $b\to u\bar c s$ amplitudes in $B^-\to D^{(*)0}K^{(*)-}$ decays~\cite{babar_gammacomb,belle_dalitz_moddep,lhcb_gammacomb}. Since the process is dominated by tree amplitudes and the hadronic parameters are extracted directly from data, the measurement of $\gamma$ through these decays is theoretically very clean~\cite{brod_zupan}. Some sources of bias have been neglected until now but might become significant at next generation $B$ physics experiments, where the experimental precision is expected to go below $1^\circ$~\cite{lhcb_upgrade,belle2}. These include the effect of mixing and $CP$ violation in $D$ decays~\cite{dmix_gamma0,dmix_gamma1,dmix_gamma2,gamma_cp} and in the neutral kaon system~\cite{grossman_savastio}.

In this paper we examine the impact of $D-\bar D$ mixing in the extraction of $\gamma$ with flavor-tagged $B\to D^{(*)0}X_s$ and $B^-\to D^{(*)0}\pi^-$ decays, with $X_s=K^{(*)}+n\pi$ ($n\geq 0$). A number of topics discussed in previous papers~\cite{dmix_gamma0,dmix_gamma1,dmix_gamma2} are further analyzed and additional aspects are investigated. In Sec.~\ref{sec:rates} we derive the rates of the decays $B^\pm\to Dh^\pm$ ($h=K,\pi$) including the effect of $D-\bar D$ mixing and without assuming $CP$ conservation in mixing or in decay. In Sec.~\ref{sec:ads}-\ref{sec:dalitz} we compute the leading corrections due to $D-\bar D$ mixing in the main analysis methods to extract the angle $\gamma$. In Sec.~\ref{sec:time_cut} we point out that the effect of charm mixing depends on how the signal selection efficiency varies as a function of the $D$ proper time and we derive the correction factor in terms of the experimental acceptance and time resolution function. The conditions under which the leading corrections become quadratic in the mixing parameters $x=\Delta m/\Gamma$ and $y=\Delta\Gamma/2\Gamma$, as opposed to linear, are discussed in Sec.~\ref{sec:quadratic}, where we point out some limitations to the applicability of this scenario. In Sec.~\ref{sec:bias} we derive the analytical formula for the bias on $\gamma$ when $D-\bar D$ mixing is ignored in the $B$ rates but not in the measurement of the $D$ decay parameters, showing that even though the bias could be in principle as large as $3^\circ$ in $B^-\to D^{(*)0}K^{(*)-}$ decays, it is found to be $\lesssim 1^\circ$ due to the measured value of the strong phase between the $b\to u\bar c s$ and $b\to c\bar u s$ amplitudes. On the other hand, the effect in $B^-\to D^{(*)0}\pi^-$ decays cannot be ignored. In Sec.~\ref{sec:btodxs} the discussion is extended to the decays $B\to D^{(*)0}X_s$ with $X_s=K^{(*)}+n\pi$ ($n\geq 0$), while in Sec.~\ref{sec:strategy} considerations on the advantage of a global combination of the $\gamma$-related observables and charm mixing measurements are presented. 

\section{$B^-\to D^0h^-$ rates ($h=\pi,K$) in the presence of $D-\bar D$ mixing}\label{sec:rates}
In this section we introduce the formalism and derive the main equations that will be used in the rest of the paper.
We define the eigenstates $|D_{1,2}\rangle$ of the effective Hamiltonian in the subspace spanned by $|D^0\rangle$ and $|\bar D^0\rangle$ as
\begin{equation}
|D_{1,2}\rangle\equiv p|D^0\rangle\pm q|\bar D^0\rangle\label{eq:d12_def},
\end{equation}
with eigenvalues $\lambda_k=m_k-\frac{i}{2}\Gamma_k$ ($k=1,2$). We adopt the convention $CP|D^0\rangle=|\bar D^0\rangle$ and $CP|\bar D^0\rangle=|D^0\rangle$, and we require that $CP|D_{1,2}\rangle=\pm|D_{1,2}\rangle$ if $CP$ is conserved. This condition implies $q/p=1$ if $CP$ is a symmetry of the Hamiltonian. We define the $D-\bar D$ mixing parameters
\begin{eqnarray}
&&m\equiv\frac{m_1+m_2}{2},\hspace{0.9cm}\Gamma\equiv\frac{\Gamma_1+\Gamma_2}{2},\\
&&x\equiv \frac{m_1-m_2}{\Gamma},\label{eq:x}\hspace{1cm}y\equiv \frac{\Gamma_1-\Gamma_2}{2\Gamma}.\label{eq:y}
\end{eqnarray}
Experimentally it is found~\cite{hfag12}
\begin{equation}
x=\left(0.63^{+0.19}_{-0.20}\right)\times 10^{-2},\hspace{0.4cm}y=(0.75\pm 0.12)\times 10^{-2}.\label{eq:xy_meas}
\end{equation}
From eq.~(\ref{eq:d12_def}) it follows
\begin{eqnarray}
&&|D^0(t)\rangle=g_+(t)|D^0\rangle+\frac{q}{p}g_-(t)|\bar D^0\rangle,\label{eq:d0_time}\\
&&|\bar D^0(t)\rangle=g_+(t)|\bar D^0\rangle+\frac{p}{q}g_-(t)|D^0\rangle,\label{eq:d0bar_time}
\end{eqnarray}
with 
\begin{equation}
g_\pm(t)\equiv\frac{e^{-i\lambda_1 t}\pm e^{-i\lambda_2 t}}{2}.\label{eq:gdef}
\end{equation}
The amplitudes of the decay $B^-\to [f]_D K^-$ and its $CP$-conjugated process can be written as
\begin{eqnarray}
&&A(B^-\to[f]_DK^-)=A_D A_f(t)+A_{\bar D}\bar A_f(t),\label{eq:bminus_amp}\\
&&A(B^+\to[\bar f]_DK^+)=\AbarDbar\Abarfbar(t)+\AbarD \Afbar(t),\label{eq:bplus_amp}
\end{eqnarray}
where $[f]_D$ represents any final state originating from the decay of $D^0$ or $\bar D^0$ and
\begin{eqnarray}
&&\AD\equiv \langle D^0K^-|H|B^-\rangle,\hspace{0.2cm}\ADbar\equiv\langle\bar D^0K^-|H|B^-\rangle,\\
&&\AbarDbar\equiv\langle \bar D^0K^+|H|B^+\rangle,\hspace{0.2cm}\AbarD\equiv\langle D^0K^+|H|B^+\rangle,\\
&&\Af\equiv\langle f|H|D^0\rangle\label{eq:af_def},\hspace{1.14cm}\Abarf\equiv\langle f|H|\bar D^0\rangle\label{eq:abarf_def},\\
&&\Af(t)\equiv g_+(t)\Af+\frac{q}{p}g_-(t)\Abarf\label{eq:af_time},\\
&&\Abarf(t)\equiv g_+(t)\Abarf+\frac{p}{q}g_-(t)\Af\label{eq:abarf_time}.
\end{eqnarray}
The time-integrated decay rate of $B^-\to [f]_DK^-$ is derived from eq.~(\ref{eq:bminus_amp}) and is proportional to
\begin{widetext}
\begin{eqnarray}
\bmampint=&&\left|\BmAmpOne\right|^2\IntGp\nonumber\\
&&+\left|\BmAmpTwo\right|^2\IntGm\nonumber\\
&&+2\Re\left[(\BmAmpOne)\left(\BmAmpTwo\right)^* \right]\IntReGpGm\nonumber\\
&&-2\Im\left[(\BmAmpOne)\left(\BmAmpTwo\right)^*\right]\IntImGpGm.\ \label{eq:Bm_master}
\end{eqnarray}
\end{widetext}
$\epsilon(t)$ is the signal selection efficiency as a function of the $D$ proper time and its effect is discussed in Sec.~\ref{sec:time_cut}. The rate of $B^+\to [\bar f]_D K^+$ is obtained from eq.~(\ref{eq:Bm_master}) with the substitution $\AD\to\AbarDbar$, $\ADbar\to\AbarD$, $\Af\to\Abarfbar$, $\Abarf\to\Afbar$, $q/p\rightarrow p/q$. 
We define $r_f$, $\delta_f$, $r_B$ and $\delta_B$ as:
\begin{eqnarray}
&&r_f e^{-i\delta_f}\equiv\frac{\Af}{\Abarf},\hspace{1cm}r_B e^{i(\delta_B-\gamma)}\equiv\frac{\ADbar}{\AD}.\label{eq:rb_def}
\end{eqnarray}
Calculating the integrals in eq.~(\ref{eq:Bm_master}) (see the Appendix) assuming a constant $\epsilon(t)$, it follows:
\begin{eqnarray}
\Gamma(B^-&&\to [f]_D K^-)\propto|\AD\Abarf|^2\bigg[\nonumber\\
&&|A_1|^2 \left(1+\frac{-x^2+y^2+2x^2y^2}{2(1+x^2)(1-y^2)}\right)\nonumber\\
&&+|A_2|^2 \frac{x^2+y^2}{2(1+x^2)(1-y^2)}\nonumber\\
&&-\Re[A_1A^*_2]\frac{y}{1-y^2}-\Im[A_1A^*_2]\frac{x}{1+x^2}\bigg],\label{eq:master_rate}
\end{eqnarray}
where
\begin{eqnarray}
&&A_1\equiv\rf+\rb,\label{eq:a1}\\
&&A_2\equiv\qp+\pq\rb\rf.\label{eq:a2}
\end{eqnarray}
The rate of $B^+\to[\bar f]_DK^+$ is derived from Eqs.~(\ref{eq:master_rate})-(\ref{eq:a2}) with the substitution $\gamma\to-\gamma$, $q/p\to p/q$ and $\rf\to\rbarfbar$, where $\rbarfbar=\rf$ if $CP$ is conserved in the $D$ decay. 

Equation~(\ref{eq:master_rate}) applies to any $D^0$ final state and any $B^-\to D^{(*)0}h^-$ decay ($h=K$, $\pi$), provided that $r_f$, $\delta_f$, $r_B$, $\delta_B$ and $\AD$ are replaced with the corresponding parameters. The general case of flavor-tagged $B\to D^{(*)0}X_s$ decays is discussed in Sec.~\ref{sec:btodxs}. 
In multibody $D$ decays $r_f$ and $\delta_f$ are functions of the position in the phase space of $f$. The impact of the $D-\bar D$ mixing corrections depends on the value of $\xyOrb$. For $B^-\to D^0K^-$ $\xyOrb\approx 0.1$~\cite{babar_gammacomb,belle_dalitz_moddep,lhcb_dalitz}. In the case of $B^-\to D^0\pi^-$, $r_{B, \pi}$ has not been measured yet but it is expected to be approximately $r_B |V_{cd}V_{us}/V_{ud}V_{cs}|\approx 0.005$, so that $\sqrt{x^2+y^2}/r_{B,\pi}\sim\mathcal{O}(1)$~\cite{dmix_gamma1}. 

In the following sections we assume $CP$ conservation in charm mixing ($q/p=1$) and decay, and we usually ignore terms of the order $\mathcal{O}(x^2+y^2)$ that can be neglected if compared to the current experimental precision. However, we will refer to the general expression of Eq.~(\ref{eq:master_rate}) in Sec.~\ref{sec:strategy} where we discuss a strategy to extract~$\gamma$. The signal selection effciency as a function of the $D$ proper time is assumed to be constant till Sec.~\ref{sec:time_cut}, where we discuss the correction factors to apply in the general case.

%
%
\section{The ADS method}\label{sec:ads}
In the Atwood-Dunietz-Soni (ADS) method~\cite{ads} the $D^0$ is reconstructed into a doubly Cabibbo-suppressed decay. In the following discussion we consider $D^0\to K^+\pi^-$ as an example. The ratio of the Cabibbo-suppressed and Cabibbo-allowed rates is measured separately for $B^+$ and $B^-$:
\begin{equation}
R^\mp_K\equiv\frac{\Gamma(B^\mp\to[K^\pm\pi^\mp]_DK^\mp)}{\Gamma(B^\mp\to[K^\mp\pi^\pm]_DK^\mp)}.\label{eq:ads_rates}
\end{equation}
Historically, the ADS observables are expressed also in terms of the $CP$ asymmetry and the average ratio, defined as $A_{ADS}\equiv (R^-_K-R^+_K)/(R^-_K+R^+_K)$ and $R_{ADS}\equiv (R^-_K+R^+_K)/2$, respectively. 
The ratios $R^\mp_\pi$ for $B^\mp\to[K^\pm\pi^\mp]_D\pi^\mp$ are defined analogously. From Eq.~(\ref{eq:master_rate}), neglecting terms quadratic in $x$ and $y$ or $\lesssim\mathcal{O}(10^{-2})$ with respect to the leading terms, the rate of the suppressed decay is 
\begin{eqnarray}
\Gamma(B^-&&\to[f]_D K^-)\propto\left|\AD\Abarf\right|^2\Big[\nonumber\\
&&r^2_f+r^2_B+2\,r_f r_B\cos(\delta_B-\gamma+\delta_f)\nonumber\\
&&-y\,r_f\cos\delta_f-y\,r_B\cos(\delta_B-\gamma)\nonumber\\
&&+x\,r_f\sin\delta_f-x\,r_B\sin(\delta_B-\gamma)\ \Big].\label{eq:ads_rate1}
\end{eqnarray}
For $f=K^+\pi^-$ it is found  $r_f\sim 0.06$ and $\delta_f\sim 200^\circ$~\cite{hfag12,hfag_convention_note}. Analogously, neglecting terms $\lesssim\mathcal{O}(10^{-4})$ compared to the main one, the rate for the Cabibbo-allowed process is
\begin{eqnarray}
\Gamma(B^-&&\to [\bar f]_D K^-)\propto\left|\AD\Abarf\right|^2\Big[\nonumber\\
&&1+r^2_f r^2_B +2\,r_f r_B\cos(\delta_B-\gamma-\delta_f)\nonumber\\
&&-y\,r_f\cos\delta_f-y\,r_B\cos(\delta_B-\gamma)\nonumber\\
&&-x\,r_f\sin\delta_f+x\,r_B\sin(\delta_B-\gamma)\ \Big],\label{eq:ads_rate5}
\end{eqnarray}
where $\Abarfbar/\Afbar=\rf$ and $|\Afbar/\Abarf|=1$ were used, which are valid when $CP$ is conserved in the decay. Using Eq.~(\ref{eq:ads_rate1}) and~(\ref{eq:ads_rate5}), and neglecting terms $\lesssim\mathcal{O}(10^{-2})$ with respect to the main ones, the ratios $R^\mp_K$ can be written as a function of the physics parameters as
\begin{eqnarray}
R^\mp_K&=& r^2_f+r^2_B+2\,r_f r_B\cos(\delta_B\mp\gamma+\delta_f) \nonumber\\
&&-y\,r_f\cos\delta_f -y\,r_B\cos(\delta_B\mp\gamma)\nonumber\\
&&+x\,r_f\sin\delta_f -x\,r_B\sin(\delta_B\mp\gamma).\label{eq:ads_rates_dmix}
\end{eqnarray}
A similar relation holds for $R^\mp_\pi$, provided that $r_B$ and $\delta_B$ are replaced with $r_{B,\pi}$ and $\delta_{B,\pi}$, respectively. It was already pointed out in Sec.~\ref{sec:rates} that the $D-\bar D$ mixing corrections are of the order of 10\% relative to the main term containing $\gamma$ in $B^-\to D^0 K^-$ ($\propto r_f\, r_Be^{i(\delta_B+\delta_f-\gamma)}$). They can sum up to $\sim 1\times 10^{-3}$, to be compared with the current experimental uncertainty on the measured ratios which is at the level of $2-3\times 10^{-3}$~\cite{lhcb_adsglw}. 
In the case of $B^-\to D^0\pi^-$ $r_{B,\pi}\sim 0.005$ is of the same order of magnitude as $x$ and $y$, and therefore the corrections are comparable in size to the terms containing $\gamma$. In this case even with the current experimental precision it is necessary to take $D-\bar D$ mixing into account. It is worth noting that the systematic uncertainty on $R^\pm_\pi$ measured by the LHCb Collaboration~\cite{lhcb_adsglw} is of the order $\mathcal{O}(x^2+y^2)\sim\mathcal{O}(r^2_{B,\pi})$, therefore the quadratic terms in Eq.~(\ref{eq:master_rate}) may become relevant in future measurements~\cite{lhcb_upgrade,belle2}.

To understand the effect of ignoring the $D-\bar D$ mixing corrections in the ADS observables it is useful to express the ratios $R^\pm_K$ in terms of the cartesian coordinates $x_{B\pm}\equiv r_B\cos(\delta_B\pm\gamma)$ and $y_{B\pm}\equiv r_B\sin(\delta_B\pm\gamma)$.
Except for terms quadratic in $x$ and $y$, Eq.~(\ref{eq:ads_rates_dmix}) can be written as
\begin{eqnarray}
R^\mp_K&=&(x_{B\mp}-y/2+r_f\cos\delta_f)^2\nonumber\\
&&+(y_{B\mp}-x/2-r_f\sin\delta_f)^2,\label{eq:ads_circ}
\end{eqnarray}
which represents two circles in the plane $(x_{B\mp},y_{B\mp})$ centered at $(-r_f\cos\delta_f-y/2,r_f\sin\delta_f-x/2)$ and with radius $\sqrt{R^\mp_K}$. Ignoring $D-\bar D$ mixing corresponds to measuring $x'_{B\pm}=\xb-y/2$ and $y'_{B\pm}=\yb-x/2$ instead of $\xb$ and $\yb$, respectively. We will come back on this point in Sec.~\ref{sec:dalitz} and when we estimate the bias of $\gamma$ in Sec.~\ref{sec:bias}.

In the above discussions we have considered $D^0\to K^+\pi^-$ as an example, but analogous results apply to other Cabibbo-suppressed decay modes such as $D^0\to K^+\pi^-\pi^0$ or $D^0\to K^+\pi^+\pi^-\pi^-$ integrated over the phase space, with the introduction of a coherence factor $\kappa$ before the terms linear in $r_f$ in Eqs.~(\ref{eq:ads_rate1}-\ref{eq:ads_rates_dmix}) (see for example~\cite{ads_kpipi0_babar,ads_k3pi_lhcb}). Equation~(\ref{eq:ads_circ}) is modified by replacing $r_f$ with $\kappa r_f$ and adding the term $1-(\kappa r_f)^2$ in the right hand-side. Therefore, the conclusion that ignoring $D-\bar D$ mixing corresponds to measuring $(\xbprime,\ybprime)$ instead of $(\xb,\yb)$ is unchanged.
 The case where a Dalitz plot analysis of the $D$ final state is performed is discussed in Sec.~\ref{sec:dalitz}.

\section{The GLW method}\label{sec:glw}
In the Gronau-London-Wyler (GLW) method~\cite{glw} the $D$ meson is reconstructed in $CP$-eigenstate final states, such as $K^+K^-$ ($CP$-even) or $K^0_S\pi^0$ ($CP$-odd). The Cabibbo-allowed decay $D^0\to K^-\pi^+$ is also reconstructed and used as normalization mode. The quantities
\begin{eqnarray}
&&R^{\bar f}_{K/\pi}\equiv\frac{\Gamma(B^-\to[\bar f]_DK^-)+\Gamma(B^+\to[f]_DK^+)}{\Gamma(B^-\to[\bar f]_D\pi^-)+\Gamma(B^+\to[f]_D\pi^+)},\ \ \label{eq:glw_ratios}\\
&&A^{\bar f}_h\equiv\frac{\Gamma(B^-\to[\bar f]_Dh^-)-\Gamma(B^+\to[f]_Dh^+)}{\Gamma(B^-\to[\bar f]_Dh^-)+\Gamma(B^+\to[f]_Dh^+)},\label{eq:glw_acp}
\end{eqnarray}
are measured, with $\bar f=f_{CP\pm}$ or $K^-\pi^+$, where $f_{CP\pm}$ indicates a generic $CP$-eigenstate state. The $CP$ asymmetries $A^{\bar f}_h$ and the double ratios $R^{f_{CP\pm}}_{K/\pi}/R^{K^-\pi^+}_{K/\pi}$ do not depend on $|\AD|$ and $|A_{D\pi}|$, and can be used to constrain $\gamma$ together with the hadronic parameters $r_B$, $\delta_B$, $r_{B,\pi}$ and $\delta_{B,\pi}$. Their measurement is experimentally advantageous because a number of uncertainties cancel out in the ratios. $A^{f_{CP\pm}}_K$ and $R^{f_{CP\pm}}_{K/\pi}/R^{K^-\pi^+}_{K/\pi}$ are often called $A_{CP\pm}$ and $R_{CP\pm}$, respectively. 

As done for the ADS method, we derive the relations between the observables and the physics parameters including the $D-\bar D$ mixing corrections. 
From Eqs.~(\ref{eq:master_rate}-\ref{eq:a2}), assuming $CP$ conservation in $D$ decay and mixing and using $\rf=\eta_\pm =\pm 1$ for $f=f_{CP\pm}$, it is $A_2=\eta_\pm A_1$ and 
\begin{eqnarray}
\Gamma(B^-&&\to [f_{CP\pm}]_D K^-)\propto|\AD\bar A_{\fcp}|^2\nonumber\\
&&\times\frac{1+r^2_B+2\eta_\pm r_B\cos(\delta_B-\gamma)}{1+\eta_\pm\,y}.\label{eq:glw_dk}
\end{eqnarray}
An analogous relation holds for $B^-\to D^0\pi^-$ with the substitution $\AD\to A_{D\pi}$, $r_B\to r_{B,\pi}$ and $\delta_B\to\delta_{B,\pi}$. The rates of the $CP$-conjugated processes are obtained with the replacement $\gamma\to-\gamma$. It is worth noting that the relation~(\ref{eq:glw_dk}) is exact, that is no quadratic or higher-order terms in $x$ and $y$ were neglected. Therefore, $R^{f_{CP\pm}}_{K/\pi}$ and $A^{f_{CP\pm}}_h$ are not affected by $D-\bar D$ mixing because the factor $1+\eta_\pm\,y$ cancels out in the ratios. This conclusion is still valid for an arbitrary selection efficiency $\epsilon(t)$ (see Sec.~\ref{sec:time_cut}) provided that it is the same for all the decays involved. The factor $1/(1+\eta_\pm\,y)$ is replaced by $\int^{+\infty}_0 |g_+(t)+\eta_\pm\,g_-(t)|^2\epsilon(t)\,dt$.
 
Charm mixing terms only appear in $R^{K^-\pi^+}_{K/\pi}$, which is defined as the ratio of the decay rates of $B^-\to D^0K^-$ and $B^-\to D^0\pi^-$ for the Cabibbo-allowed $D^0$ mode. The rate of $B^-\to D^0K^-$ is given by Eq.~(\ref{eq:ads_rate5}) and a similar relation holds for $B^-\to D^0\pi^-$ with the usual replacement $r_B\to r_{B,\pi}$, $\delta_B\to\delta_{B,\pi}$ and $\AD\to A_{D\pi}$. The resulting $D-\bar D$ mixing terms partially cancel out in $R^{K\pi}_{K/\pi}$ and are $\lesssim\mathcal{O}(10^{-3})$ compared to the main term. 

To summarize, the $D-\bar D$ mixing corrections in the GLW method cancel out in the $CP$ asymmetries and are of the order $\mathcal{O}(r_B\sqrt{x^2+y^2})$ in $R^{f_{CP\pm}}_{K/\pi}/R^{K^-\pi^+}_{K/\pi}$. This conclusion differs from the one in~\cite{dmix_gamma1} because at that time it was probably not clear that it is experimentally advantageous to express the GLW observables in terms of $CP$ asymmetries and double ratios.

\section{The Dalitz method}\label{sec:dalitz}
In the Dalitz method the $D$ meson is reconstructed in a 3-body final state such as $D^0\to K^0_S\pi^+\pi^-$ [Giri-Grossman-Soffer-Zupan (GGSZ) method]~\cite{ggsz,bondar_dalitz} or $K^+\pi^-\pi^0$~\cite{ads}. The use of 4-body decays has also been investigated~\cite{rademacker_wilkinson}. In the following we consider the decay $D^0\to K^0_S h^+h^-$ ($h=\pi,K$), but similar conclusions apply to other decay modes. First we analyze the model-dependent analysis, then we will comment on the model-independent approach.

 Assuming no $CP$ violation, we write the decay amplitudes of $D^0$ and $\bar{D}^0$ as $A_f\equiv f_-=f(m^2_-,m^2_+)$ and $\bar{A}_f\equiv f_+=f(m^2_+,m^2_-)$, where $m^2_-$ and $m^2_+$ are the squared masses of $K^0_Sh^-$ and $K^0_Sh^+$, respectively. From Eq.~(\ref{eq:master_rate}), neglecting terms quadratic in $x$, $y$ and introducing the cartesian coordinates $x_{B\pm}=r_B\cos(\delta_B\pm\gamma)$, $y_{B\pm}=r_B\sin(\delta_B\pm\gamma)$, the yield is:
\begin{eqnarray}
\Gamma(B^-&&\to[f]_D K^-)\propto\nonumber\\
&&|f_-|^2(1-x_{B-}\,y+y_{B-}\,x)\nonumber\\
&&+|f_+|^2\left(r^2_B-x_{B-}\,y-y_{B-}\,x\right)\nonumber\\
&&+2\Re[f_-f^*_+]\left(x_{B-}-\frac{y}{2}(1+r^2_B)\right)\nonumber\\
&&+2\Im[f_-f^*_+]\left(y_{B-}-\frac{x}{2}(1-r^2_B)\right).\label{eq:dalitz_dmix1}
\end{eqnarray}
Introducing the shifted coordinates $x'_{B\pm}=x_{B\pm}-y/2$ and $y'_{B\pm}=y_{B\pm}-x/2$ as in Sec.~\ref{sec:ads}, Eq.~(\ref{eq:dalitz_dmix1}) can be written as
\begin{eqnarray}
&&\Gamma(B^-\to [f]_DK^-)\propto\nonumber\\
&&|f_-|^2+r'^2_{B-}|f_+|^2+2x'_{B-}\Re[f_-f^*_+]+2y'_{B-}\Im[f_-f^*_+],\label{eq:dalitz_dmix3}\nonumber\\
\end{eqnarray}
where $r'^2_{B\mp}=x'^2_{B\mp}+y'^2_{B\mp}$ and terms $\lesssim\mathcal{O}(10^{-2})$ with respect to $x$ and $y$ have been neglected. In Eq.~(\ref{eq:dalitz_dmix3}) the term $(1-x_{B-}\,y+y_{B-}\,x)$ multiplying $|f_-|^2$ was factored out. In any case in the region $|f_-/f_+|^2\lesssim 1$, where the sensitivity to $\gamma$ is larger, the quantity $(-x_{B-}\,y+y_{B-}\,x)|f_-|^2$ is suppressed by a factor $\lesssim\mathcal{O}(r_B)$ compared to the leading $D-\bar D$ mixing terms. Equation~(\ref{eq:dalitz_dmix3}) and its $CP$-conjugated version are the equations used to extract the cartesian coordinates with the model-dependent GGSZ analysis in the absence of $D-\bar D$ mixing, with $(x_{B\mp},y_{B\mp})$ replaced by $(x'_{B\mp},y'_{B\mp})$. Therefore, if the amplitudes $f_\pm$ are correctly measured without neglecting linear $D-\bar D$ mixing effects (as for example in the time-dependent Dalitz plot analysis to measure $x,y$ from $D^0\to K^0_S\pi^+\pi^-$ decays~\cite{babar_dalitz_dmix}, or in time-integrated $\Psi(3770)\to D\bar D$ decays at charm factories~\cite{psi3770_conditions}), then ignoring charm mixing in the measurement of $\gamma$ corresponds to measuring $(\xbprime,\ybprime)$ instead of $(\xb,\yb)$. This is the case of the BaBar measurement~\cite{babar_gammacomb}, where the nominal Dalitz models are determined in a time-dependent $D-\bar D$ mixing measurement~\cite{babar_dalitz_dmix}. Anyway, the bias in $(\xb,\yb)$ is about one order of magnitude smaller than the statistical error and comparable to the systematic uncertainty associated to the Dalitz model, which might be difficult to improve and at present is considered as a potential irreducible limitation of the model-dependent method. In the model-dependent Dalitz plot measurement of $\gamma$ performed by Belle~\cite{belle_dalitz_moddep} the $D-\bar D$ mixing was ignored both in the $B$ rates and in the extraction of $f_\pm$ from flavor-tagged, time-integrated $D^0\to K^0_S \pi^+\pi^-$ decays. In this case the bias in the extraction of $(\xb,\yb)$ depends on the form of $f_\pm$ and its precise estimate requires a simulation of the measurement, though in general the magnitude is reduced compared to the case where $D-\bar D$ mixing is not ignored in the extraction of $f_\pm$.

The model-independent approach~\cite{ggsz} is free from the uncertainty associated to the Dalitz model description but it relies on the measurement of a set of hadronic parameters at charm factories~\cite{cleoc_modind}. The Dalitz plot is divided into $2N$ bins chosen to be symmetric under the exchange $m^2_+\leftrightarrow m^2_-$, and the number of $B^\mp\to [K^0_S\pi^+\pi^-]_DK^\mp$ decays is measured in each bin. The main relation can be obtained from Eq.~(\ref{eq:dalitz_dmix1}) with the substitution $|f_-|^2\to K_j$, $|f_+|^2\to K_{-j}$, $f_-f^*_+\to\sqrt{K_jK_{-j}}(C_j+i S_j)$, where the index $j$ indicates the $j^{th}$ Dalitz plot bin and $K_j$ is proportional to $|A(D^0\to K^0_S\pi^+\pi^-)|^2$ integrated over bin $j$ of the Dalitz plot. The parameters $C_j$ and $S_j$ contain information on the strong phase difference between $A(D^0\to K^0_S\pi^+\pi^-)$ and $A(\bar D^0\to K^0_S\pi^+\pi^-)$ in each bin $j$, and are measured at charm threshold~\cite{cleoc_modind}. From Eq.~(\ref{eq:dalitz_dmix3}) the number of $B^\mp$ decays in each bin $j$ is
\begin{eqnarray}
N^\mp_j\propto K_j+r'^2_{B\mp} K_{-j}+2\sqrt{K_jK_{-j}}\left(x'_{B\mp}C_j+y'_{B\mp}S_j\right).\label{eq:dalitz_modind}\nonumber\\
\end{eqnarray}
The result is analogous to the model-dependent case. If $K_j$, $C_j$ and $S_j$ are measured without ignoring charm mixing, then ignoring it in the measurement of the $B^\pm$ rates corresponds to measuring $(\xbprime,\ybprime)$ instead of $(\xb,\yb)$. On the other hand, it has been shown in~\cite{bondar_etal_modind_dmix} that ignoring $D-\bar D$ mixing at all stages of the analysis introduces a bias $\lesssim 0.2^\circ$ in the extraction of $\gamma$. In this regard see also the discussion in Sec.~\ref{sec:quadratic}.

In the case of $B^-\to D^0\pi^-$ the bias is comparable to the magnitude of the cartesian coordinates since $r_{B,\pi}\lesssim\mathcal{O}(0.01)$. Therefore, either the model-independent approach ignoring $D-\bar D$ mixing at all stages of the analysis should be used~\cite{bondar_etal_modind_dmix}, or the charm mixing corrections should be included. No measurement of $\gamma$ using the Dalitz method with $B^-\to D^{(*)0}\pi^-$ decays has been attempted so far. 

\section{Effect of a non-uniform signal selection efficiency as a function of the $D$ proper time}\label{sec:time_cut} 
In deriving Eq.~(\ref{eq:master_rate}) from Eq.~(\ref{eq:Bm_master}) it was assumed that the signal selection efficiency $\epsilon(t)$ as a function of the $D$ proper time is constant. However, if this assumption is not valid (see for example~\cite{lhcb_adsglw}) a correction might be required. $\epsilon(t)$ can be derived from the signal acceptance $A(t')$ as a function of the reconstructed proper time and from the time resolution function $R(t'-t)$, $\epsilon(t)=\int^{+\infty}_{-\infty}R(t'-t)A(t')dt'$.
The form of Eq.~(\ref{eq:master_rate}) still holds for a generic $\epsilon(t)$ provided that the last three terms are multiplied by $\int^{+\infty}_0 f(t)\epsilon(t) dt/(I\int^{+\infty}_0 f(t) dt)$, where $f(t)=|g_-(t)|^2$, $\Re[g_+(t)g^*_-(t)]$ and $\Im[g_+(t)g^*_-(t)]$, respectively, and $I\equiv\int^{+\infty}_0 |g_+(t)|^2\epsilon(t)dt/\int^{+\infty}_0 |g_+(t)|^2dt$.
Neglecting differences of the order $\mathcal{O}(x^2+y^2)$, the coefficients multiplying the terms linear in $x$ and $y$ have the same value
\begin{eqnarray}
\alpha&\equiv&\frac{\int^{+\infty}_0 \Re[g_+(t)g^*_-(t)]\epsilon(t) dt}{I\int^{+\infty}_0 \Re[g_+(t)g^*_-(t)]dt}\nonumber\\
&=&\frac{\int^{+\infty}_0 \Im[g_+(t)g^*_-(t)]\epsilon(t) dt}{I\int^{+\infty}_0 \Im[g_+(t)g^*_-(t)]dt}.\label{eq:alpha}
\end{eqnarray}

\begin{figure}
\includegraphics[width=8cm]{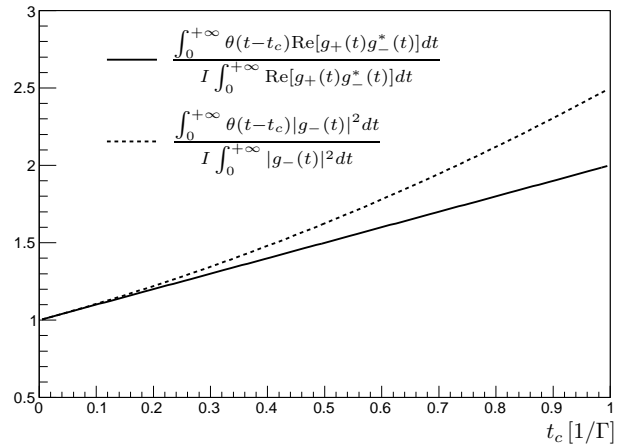}
\caption{Correction factors for the acceptance function $A(t')=\theta(t'-t_c)$ and perfect time resolution.}
\label{fig:alpha_tc}
\end{figure}

We analyze a simplified scenario where the decays with proper time between 0 and $t_c$ do not pass the selection, with $\Gamma t_c\lesssim 1$. This corresponds to assuming the acceptance function $A(t')=\theta(t'-t_c)$ and perfect time resolution, where $\theta(t')$ is the Heavyside step function. We refer to Eq.~(\ref{eq:gpgp_gmgm_timecut}) and~(\ref{eq:gpgm_timecut}) in the Appendix. Apart from the factor $e^{-\Gamma t_c}$, which is common to all integrals, from Eq.~(\ref{eq:gpgp_gmgm_timecut}) two new contributions appear, $y\sinh(y\Gamma t_c)\lesssim y^2$ and $x\sin(\Gamma t_c x)\lesssim x^2$, resulting in the additional terms $(y^2\mp x^2)(\Gamma t_c+(\Gamma t_c)^2/2)$ in $\int|g_\pm(t)|^2dt$. On the other hand, in Eq.~(\ref{eq:gpgm_timecut}) the leading effect is the multiplication of the terms linear in $x$ and $y$ by $\alpha=1+\Gamma t_c$: if the experimental precision is good enough to discriminate $x$ and $y$ with a relative precision of the order $\Gamma t_c$, the correction is not negligible. Figure~\ref{fig:alpha_tc} shows how the correction factors scale as a function of $t_c$ in the example just discussed. 

The effect can be particularly relevant for $B^-\to D^0\pi^-$~\cite{lhcb_gammacomb}\cite{lhcb_adsglw}, where $x$ and $y$ are of the same order of magnitude as $r_{B,\pi}$. In case the corrections associated to a non-uniform $\epsilon(t)$ are found to be non-negligible they should be quoted to allow their use in independent computations of $\gamma$~\cite{utfit,ckmfitter}.
 
In Secs.~\ref{sec:ads}-\ref{sec:dalitz} it is implicitly assumed that all terms linear in $x$ and $y$ are multiplied by $\alpha$, although we set it to 1 to simplify the form of the equations.  
\vspace{0.0cm}
\section{Conditions under which the leading $D-\bar D$ mixing corrections are quadratic in $x$ and $y$}\label{sec:quadratic}
It is noted in~\cite{dmix_gamma2} that the leading $D-\bar D$ mixing effects are quadratic in $x$ and $y$, as opposed to linear, when some specific conditions are satisfied. 
We briefly examine these conditions and we point out some practical limitations when they are applied to the ADS method.
Following~\cite{dmix_gamma2} and using Eq.~(\ref{eq:bminus_amp}), the rate of $B^-\to D^0K^-$ can be written as
\begin{eqnarray}
&&\Gamma(B^-\to [f]_DK^-)\propto\nonumber\\
&&|\AD|^2\left(\Gamma_f+r^2_B\bar\Gamma_f+2r_B\Re\hspace{-0.5mm}\left[e^{i(\delta_B-\gamma)}\sqrt{\Gamma_f\bar\Gamma_f}e^{i\bar\delta_f}e^{-\epsilon_f}\right]\right),\label{eq:ads_quadratic}\nonumber\\
\end{eqnarray}
where $\Gamma_f\equiv\int^{+\infty}_0|A_f(t)|^2 dt$, $\bar \Gamma_f\equiv\int^{+\infty}_0|\bar A_f(t)|^2 dt$, $\bar\delta_f$ is a strong phase and $\epsilon_f\equiv 1/8(x^2+y^2)(1/r^2_f+r^2_f)-1/4(x^2\cos 2\delta_f+y^2\sin2\delta_f)$. The terms linear in $x$ and $y$ are absorbed in $\Gamma_f$, $\bar\Gamma_f$ and $\bar\delta_f$. In the limit where $\epsilon_f=0$ Eq.~(\ref{eq:ads_quadratic}) has the same form as the rate in Eq.~(\ref{eq:master_rate}) with $x=y=0$, provided that $r^2_f$ is replaced with $\Gamma_f/\bar\Gamma_f$ and $\delta_f$ with $\bar\delta_f$. Therefore, when ($CP$-conserving) $D-\bar D$ mixing is ignored in both the $B$ and $D$ decay amplitudes, the error in the extracted value of $\gamma$ is of second order in $x/r_f$ and $y/r_f$~\cite{dmix_gamma2}.

We indicate two possible experimental drawbacks of this approach in the ADS method. The first is that what is measured from the combined $D-\bar D$ mixing measurements~\cite{hfag12} is $r^2_f$, not $\Gamma_f/\bar\Gamma_f$. $\Gamma_f/\bar\Gamma_f$ could be measured for this purpose from samples of flavor-tagged $D^0\to f$ and $\bar D^0\to f$ decays, although some care would be required because the selection should reflect the one in $B^\pm$ decays: indicating with $\alpha_1$ and $\alpha_2$ the correction factors defined in Eq.~(\ref{eq:alpha}) for the two selections, terms linear in $x$ and $y$ weighted by $\alpha_1-\alpha_2$ survive. The second drawback is that the phase $\bar\delta_f$, which can be shown to differ from $\delta_f$ by terms $\mathcal{O}(\sqrt{x^2+y^2}/r_f)$, should be obtained directly from the fit to the $B^\pm$ decay rates with a consequent reduction of the sensitivity to $\gamma$. We remark that when the approach in~\cite{dmix_gamma2} was proposed the constraints on the size of $x$, $y$ and $\delta_f$ ($f=K^+\pi^-$) were very loose or not available, and therefore a precise estimate or direct correction of $D-\bar D$ mixing effects was not possible.

We conclude that in general the corrections due to $D-\bar D$ mixing are linear and of the order $\mathcal{O}(\sqrt{x^2+y^2}/r_B)$ compared to the terms containing $\gamma$. A significant exception is represented by the GLW method discussed in Sec.~\ref{sec:glw}, but also by the model-independent Dalitz method when charm mixing is ignored at all stages of the analysis~\cite{bondar_etal_modind_dmix}. It is worth noting that the estimate of the maximum bias $|\Delta\gamma|\sim 0.2^o$ in~\cite{bondar_etal_modind_dmix} assumes the same selection efficiency $\epsilon(t)$ in all relevant decays. Otherwise new terms linear in $x$ and $y$ would survive as previously discussed for the ADS method, leading to a possible increase of the bias. 
 
 Nonetheless, we show in Sec.~\ref{sec:bias} that the bias on $\gamma$ introduced when the linear corrections are ignored is limited in $B^-\to D^{(*)0}K^{(*)-}$ decays due to the particular values of the strong phases $\delta_B$. In the case of $B^-\to D^{(*)0}\pi^-$, however, such corrections cannot be neglected.

\section{Bias in the extraction of $\gamma$ when $D-\bar D$ mixing is ignored in the $B^\pm$ rates}\label{sec:bias}
\begin{figure}[bh!]
\includegraphics[width=9cm]{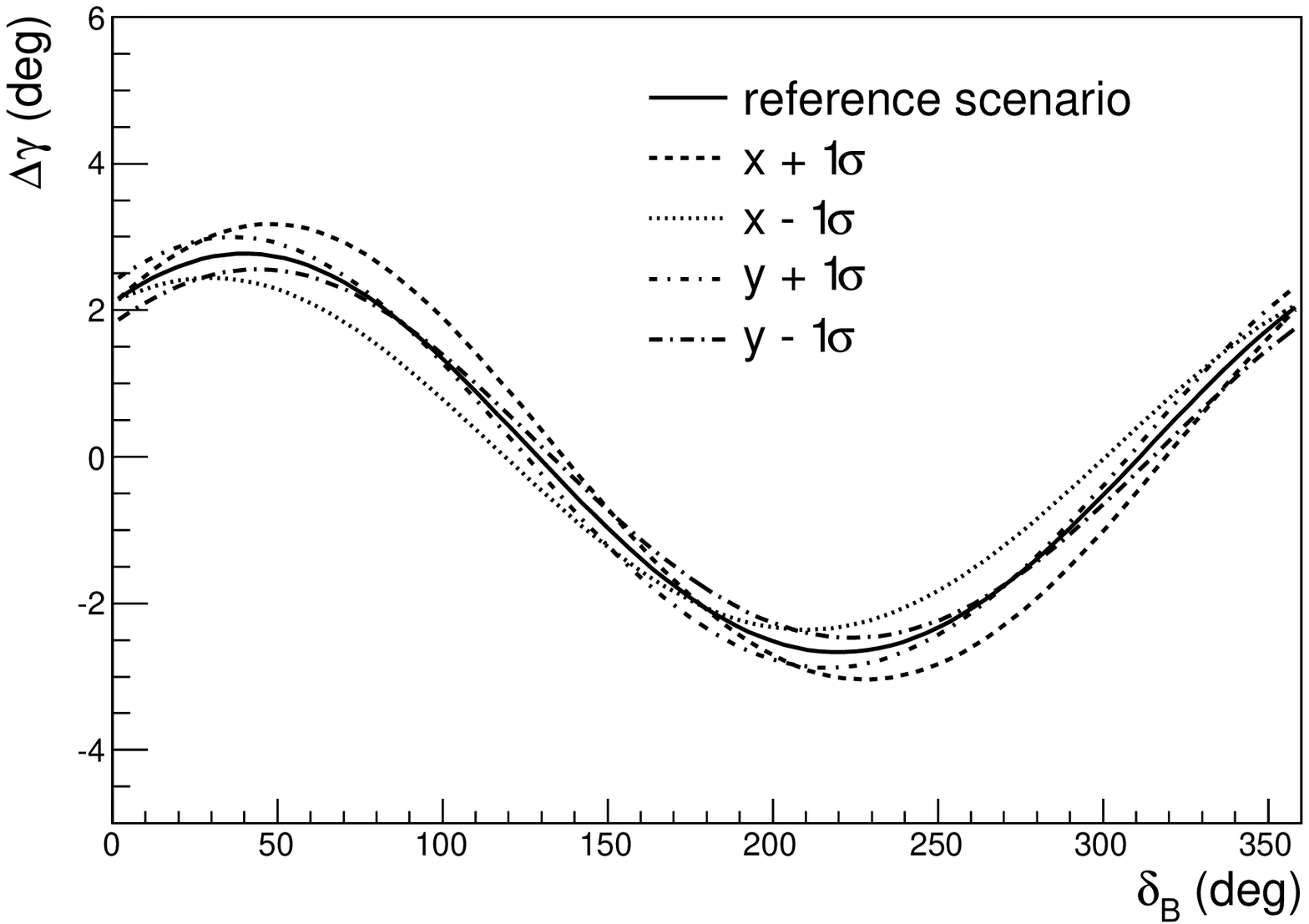}
\includegraphics[width=9cm]{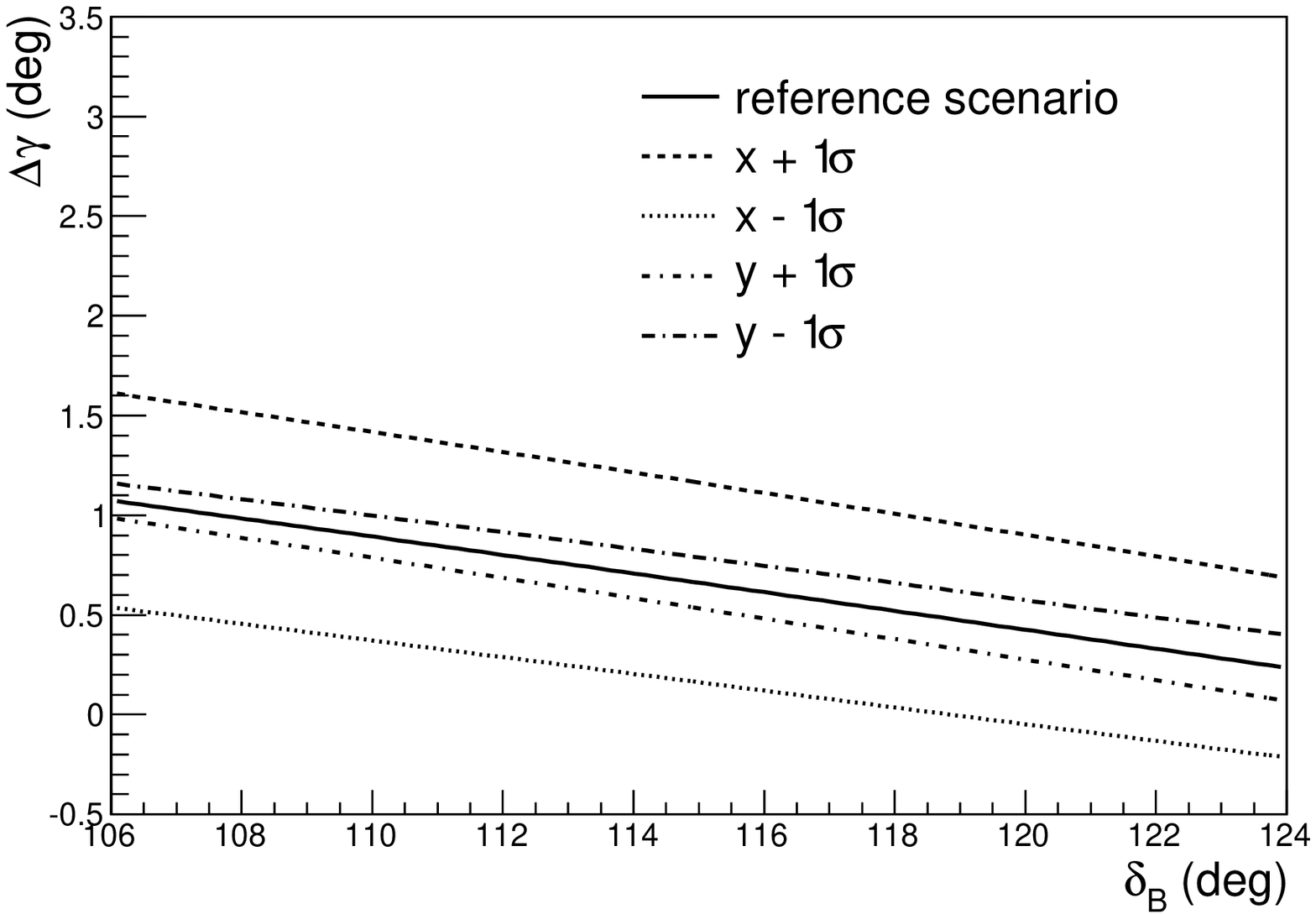}
\caption{Top plot: $\Delta\gamma$ as a function of $\delta_B$ in $B^-\to D^0K^-$ decays using the ADS and Dalitz methods when $D-\bar D$ mixing is ignored in the $B^\pm$ rates but not in the determination of the $D$ decay parameters, for different values of $x$ and $y$ (Sec.~\ref{sec:bias}). Bottom plot: $\Delta\gamma$ in the region $\delta_B=(115\pm 9)^\circ$. The resulting bias is $\Delta\gamma=(0.7\pm 0.7)^\circ$.}
\label{fig:delta_gamma1}
\end{figure}
\begin{figure*}
\includegraphics[width=7.5cm]{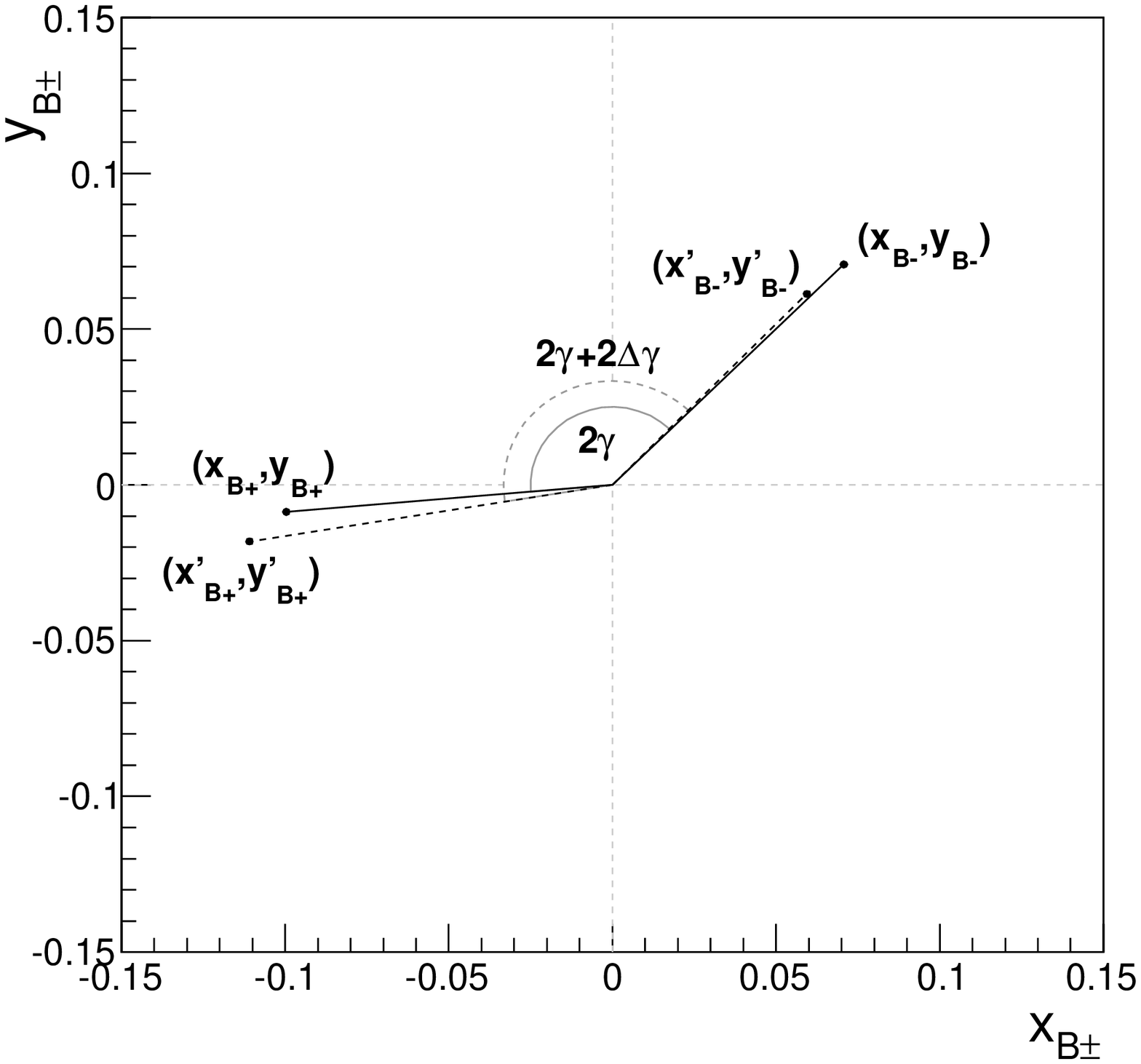}
\includegraphics[width=7.5cm]{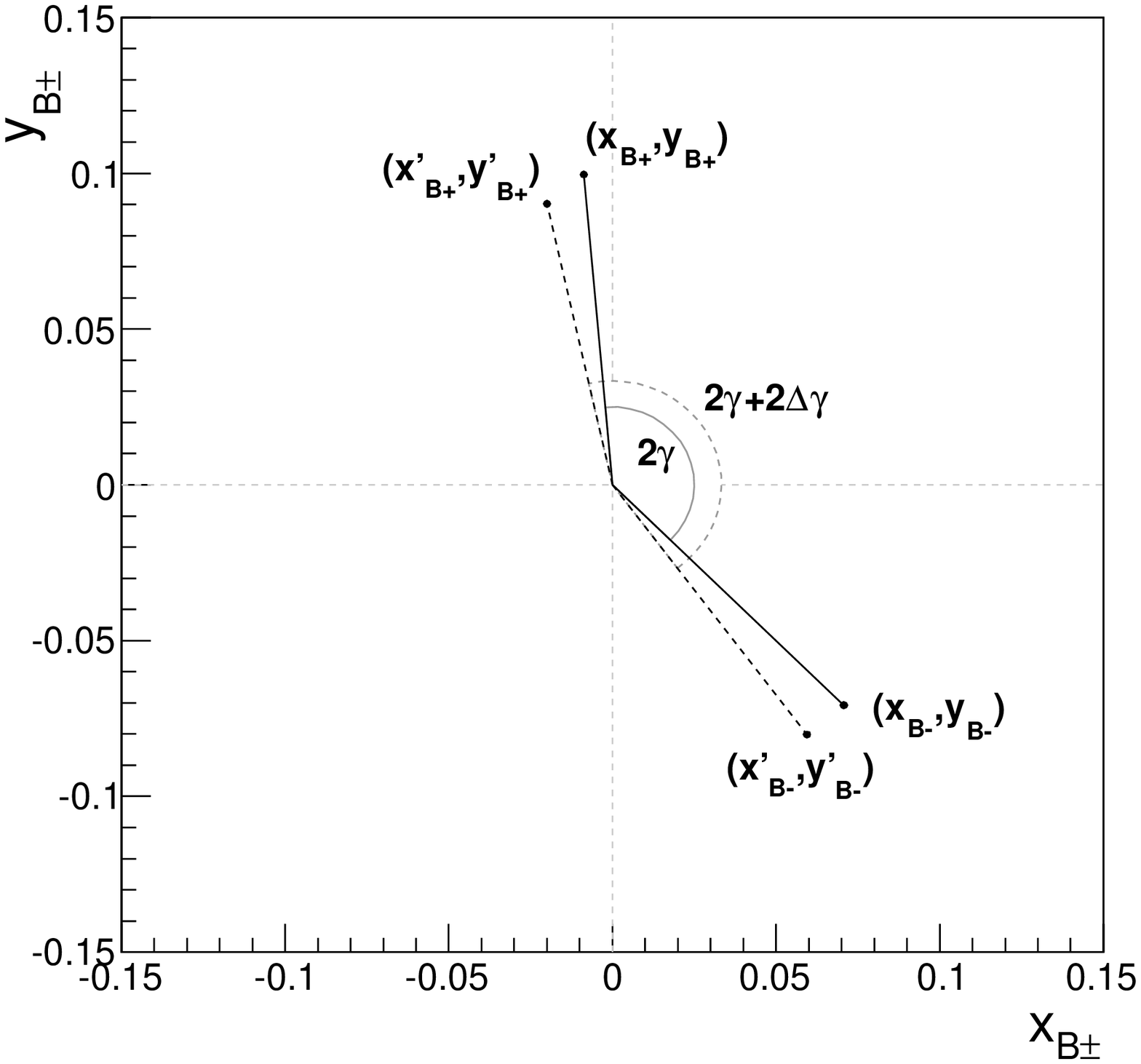}
\caption{Left plot: Geometric representation of ($\xb$,$\yb$), ($x'_{B\pm}$,$y'_{B\pm}$) and $\Delta\gamma$ assuming $r_B=0.1$, $\gamma=70^\circ$ and $\delta_B=115^\circ$. For illustration purpose the central values of $x$ and $y$ have been enlarged by a factor 3 with respect to the values in Eq.~(\ref{eq:xy_meas}). Right plot: The same configuration except for $\delta_B$, which was shifted by $-90^\circ$. In this case $|\Delta\gamma|$ is visibly larger.}
\label{fig:delta_gamma2}
\end{figure*}
Following the discussion in Sec.~\ref{sec:ads} and~\ref{sec:dalitz}, we estimate the bias in the extraction of $\gamma$ and $r_B$ using $B^-\to D^{(*)0}K^-$ decays when $D-\bar D$ mixing is ignored in the $B^\pm$ rates but not in the measurement of the $D$ decay parameters, that is assuming that $(\xbprime,\ybprime)$ are measured instead of $(\xb,\yb)$. The angle between the lines connecting ($x_{B+}$,$y_{B+}$) and ($x_{B-}$,$y_{B-}$) with $(0,0)$ is $2\gamma$. Analogously, we define $2\gamma'$ as the angle computed when $(\xb,\yb)$ are replaced with $(\xbprime,\ybprime)$. The bias $\Delta\gamma$ is defined as $\Delta\gamma=\gamma'-\gamma$. It can be shown that the following relation holds
\begin{equation}
\Delta\gamma\,(\text{rad})=\frac{\alpha}{2}\frac{\sqrt{x^2+y^2}}{r_B}\sin\gamma\,\sin(\delta_B+\delta_0),\label{eq:deltagamma_btodk}
\end{equation}
where $\delta_0=\arctan(y/x)$ and $\alpha$ is the correction factor defined in Eq.~(\ref{eq:alpha}). Equation~(\ref{eq:deltagamma_btodk}) neglects terms of the order $\mathcal{O}((x^2+y^2)^{3/2}/r^3_B)$, therefore it is accurate for $B^-\to D^{(*)0}K^-$ but should not be used for $B^-\to D^{(*)0}\pi^-$. Figure~\ref{fig:delta_gamma1}, top plot, shows how $\Delta\gamma$ varies as a function of $\delta_B$ assuming  $\gamma=70^\circ$, $r_B=0.1$, $y=0.75\times 10^{-2}$, $x=0.63\times 10^{-2}$ and $\alpha=1$. The scenarios where the value of $x$ or $y$ is changed by $\pm 1\sigma$ according to Eq.~(\ref{eq:xy_meas}) are superimposed. The bias ranges between $+3^\circ$ and $-3^\circ$ depending on the value of $\delta_B$. However, if we restrict ourselves to the measured range for $B^-\to D^0K^-$, $\delta_B=(115\pm 9)^\circ$~\cite{ckmfitter}, the bias reduces to $\Delta\gamma=(0.7\pm 0.7)^\circ$, where the error includes the uncertainty on $\delta_B$, $x$, $y$, $r_B$ and $\gamma$, and is dominated by the first two contributions. This is shown in Fig.~\ref{fig:delta_gamma1}, bottom plot. 
The bias variation as a function of $\delta_B$ is explained geometrically in Fig.~\ref{fig:delta_gamma2}. The left plot shows the position of $\xb$ and $\yb$ assuming $\gamma=70^\circ$, $\delta_B=115^\circ$ and $r_B=0.1$, together with the shifted points $(x'_{B\pm}, y'_{B\pm})$. The central values of $x$ and $y$ have been enlarged three times with respect to Eq.~(\ref{eq:xy_meas}) to make the bias more visible. In the right plot the same comparison is shown except for the value of $\delta_B$, which is shifted by $-90^\circ$. In this case $\Delta\gamma$ is visibly larger. In conclusion, the shift of $\gamma$ in the ADS and Dalitz methods when $D-\bar D$ mixing is ignored in the $B$ rates but not in the determination of the $D$ decay parameters could be in principle as large as $\sim 3^\circ$ in $B^-\to D^0K^-$ decays, but in practice it is reduced to $(0.7\pm 0.7)^\circ$ due to the particular value of $\delta_B$. Similar considerations and conclusions apply to $B^-\to D^{*0}K^-$ and $B^-\to D^0K^{*-}$. For the latter, a correction may be required to deal with the non-negligible natural width of the $K^{*-}$ and the consequent interference with other states, as discussed in Sec.~\ref{sec:btodxs}. The resulting shifts are summarized in tab.~\ref{tab:bias_table}.

With ($\xbprime$,$\ybprime$) in general the relation $r'^2_{B+}\equiv x'^2_{B+}+y'^2_{B+}=r'^2_{B-}\equiv x'^2_{B-}+y'^2_{B-}$ does not hold: $r'_{B\pm}\approx r_B\left(1-y\,\xb/(2r^2_B)-x\,\yb/(2r^2_B)\right)$. Using the measured values of $x$, $y$, $r_B$, $\gamma$ and $\delta_B$ we find for $B^-\to D^0K^-$ $r'_{B+}=(1.04\pm 0.01)\,r_B$ and  $r'_{B-}=(0.95\pm 0.01)\,r_B$. 

\begin{table}
\caption{\label{tab:bias_table}World average measurements~\cite{ckmfitter} of $r_B$ and $\delta_B$ for $B^-\to D^{(*)0}K^{(*)-}$ and resulting bias $\Delta\gamma$ when ($x'_{B\pm}=x_{B\pm}-y/2$,$y'_{B\pm}=y_{B\pm}-x/2$) are used instead of $(\xb,\yb)$. The bias for $B^-\to D^{*0}K^-$ refers to $D^{*0}\to D^0\pi^0$. For $D^{*0}\to D^0\gamma$ the bias has opposite sign due to the effective $180^\circ$ shift of $\delta_B$~\cite{bondar_gershon}. In the case of $B^-\to D^0K^{*-}$ $r_B$ is in fact $\kappa_Br_B$ (see also the discussion in Sec.~\ref{sec:btodxs}).}
\begin{ruledtabular}
\begin{tabular}{lccc}
  & $r_B$ & $\delta_{B}$ (deg) & $\Delta\gamma$ (deg) \\
\hline
$B^-\to D^0K^-$ & $0.096\pm 0.006$  & $115\pm 9$ & $0.7\pm 0.7$ \\
$B^-\to D^{*0}K^-$ & $0.121\pm 0.019$  & $-55\pm 14$ & $-0.2\pm 0.6$ \\
$B^-\to D^0K^{*-}$ & $0.140\pm 0.046$   & $110^{+31}_{-42}$ & $0.6\pm 1.1$ \\
\end{tabular}
\end{ruledtabular}
\end{table}

\section{The decays $B\to D^{(*)0}X_s$}\label{sec:btodxs}
The equations giving the yield as a function of the physics parameters derived in the previous sections are valid for any flavor-tagged $B\to D^{(*)0}X_s$ decay with $X_s=K^{(*)}+n\pi$ ($n\geq 0$) provided that the terms linear in $r_B$ are multiplied by a coherence factor $\kappa_B$, where $0\leq\kappa_B\leq 1$~\cite{gronau_plb557_198}. The values of $r_B$, $\delta_B$ and $\kappa_B$ depend on $X_s$ and its selected phase space region. 

The effect of $D-\bar D$ mixing is unchanged in the GLW method, where the factor $1/(1+\eta_\pm y)$ still multiplies the rates as in Eq.~(\ref{eq:glw_dk}). On the other hand, in the ADS and Dalitz methods when $\kappa_B\neq 1$ it is no more possible to find a transformation of the cartesian coordinates that preserves the form of the yields as a function of the physics parameters as in Eqs.~(\ref{eq:ads_circ}), (\ref{eq:dalitz_dmix3}) and~(\ref{eq:dalitz_modind}). Using the transformation $\kappa_B x''_{B\pm}\equiv \kappa_B x_{B\pm}-y/2$ and $\kappa_B y''_{B\pm}\equiv \kappa_B y_{B\pm}-x/2$, and defining $r''^2_{B\pm}\equiv x''^2_{B\pm}+y''^2_{B\pm}$, the rate of $B^-\to [f]_DX^-_s$ in the Dalitz method can be written as
\begin{eqnarray}
&&\Gamma(B^-\to [f]_D X_s^-)\propto|f_-|^2+r''^2_{B-}|f_+|^2\nonumber\\
&&+2\kappa_B\,x''_{B-}\Re[f_-f^*_+]+2\kappa_B\,y''_{B-}\Im[f_-f^*_+]+\Delta_-,\ \label{eq:dalitz_dmix4}
\end{eqnarray}
where $\Delta_\mp\equiv (1/\kappa_B-\kappa_B)(x_{B\mp}\,y+y_{B\mp}\,x)|f_+|^2$. However, even for values of $\kappa_B$ such that $(1/\kappa_B-\kappa_B)\sim\mathcal{O}(1)$, the impact of $\Delta_\mp$ on the measurement of $\kappa_B x''_{B\pm}$ and $\kappa_B y''_{B\pm}$ is expected to be small. A different definition of $(x''_{B\pm},y''_{B\pm})$ would introduce terms proportional to $\Re[f_-f^*_+]$ and $\Im[f_-f^*_+]$ in $\Delta_\mp$, with a probable increase of the bias in the extraction of $(\kappa_B x''_{B\pm},\kappa_B y''_{B\pm})$.

As a consequence, the estimate of the bias $\Delta\gamma$ when $D-\bar D$ mixing is ignored in $B$ rates but not in the measurement of the $D$ amplitudes, given in Eq.~(\ref{eq:deltagamma_btodk}) for $B^-\to D^{(*)0}K^-$ decays, can be approximately applied to a generic flavor-tagged $B\to D^{(*)0}X_s$ decay provided that $r_B$ is replaced with $\kappa_Br_B$.

\section{Combined charm mixing and $\gamma$ measurement}\label{sec:strategy}
In the previous sections we have discussed the leading corrections due to $D-\bar D$ mixing on a number of $\gamma$-related observables. Such corrections should be taken into account in the global combination to extract $\gamma$~\cite{lhcb_gammacomb,utfit,ckmfitter} when they are significant compared to the experimental uncertainties: this is the case for $B^-\to D^0\pi^-$ and might be soon the case for the ADS measurement of $B^-\to D^0[K^+\pi^-]_D K^-$~\cite{hfag12,lhcb_adsglw}. 

In our discussions we have assumed $CP$ conservation in $D-\bar D$ mixing and decay and neglected next-to-leading order terms, but in general these restrictions can be released using Eq.~(\ref{eq:master_rate}). 
A global fit may be performed to simultaneously extract the $D-\bar D$ mixing and $\gamma$-related parameters by combining all the relevant measurements. The fit can be done imposing $CP$ conservation in $D$ decay, or allowing for $CP$ violation in charm mixing and decay. The advantage of a global combination is twofold.
The measurement of $\gamma$ would exploit the full knowledge of the charm mixing parameters and $CP$ asymmetry constraints with the correlations properly taken into account. On the other hand, the charm mixing measurements could take advantage of additional constraints from $\gamma$-related observables such as the one on the strong phase difference $\delta_{K\pi}$ from the ADS method, which is necessary to exploit the knowledge of $y\prime\equiv y\cos\delta_{K\pi}-x\sin\delta_{K\pi}$~\cite{hfag12}. The constraining power on $\delta_{K\pi}$ from the current ADS measurements~\cite{hfag12,lhcb_adsglw} is comparable to the direct measurement performed by the CLEO Collaboration at the charm threshold~\cite{cleoc_deltakpi}.

\section{Conclusions}
We have examined the impact of $D-\bar D$ mixing in the rates of flavor-tagged $B\to D^{(*)0}X_s$ and $B^-\to D^{(*)0}\pi^-$ decays, with $X_s=K^{(*)}+n\pi$ ($n\geq 0$). We have computed the leading corrections, linear in the mixing parameters $x$ and $y$, for the ADS, GLW and Dalitz methods. In the GLW method the effect cancels out in the $CP$ asymmetries and is suppressed in the double ratio $R^{f_{CP\pm}}_{K/\pi}/R^{K^-\pi^+}_{K/\pi}$. 

We have observed that the effect depends on how the signal selection efficiency varies as a function of the $D$ proper time and we have estimated the correction factor in a simplified case.

We have shown that ignoring $D-\bar D$ mixing in the extraction of both the $D$ and $B$ decay amplitudes, which makes the leading corrections quadratic in $x$ and $y$ as opposed to linear, does not allow to fully exploit the available information in the ADS method. 

When $D-\bar D$ mixing is ignored in the $B^\pm$ rates but not in the measurement of the $D$ amplitude parameters, the effect in the ADS and Dalitz methods can be described at leading order by replacing the cartesian coordinates ($\xb$,$\yb$) with ($\xb-y/2$,$\yb-x/2$). We have estimated the bias $\Delta\gamma$  when these corrections are ignored in $\btodstkst$ decays finding $|\Delta\gamma|\lesssim 1^\circ$, limited by the value of the strong phases $\delta_B$. On the other hand the effect of $D-\bar D$ mixing in $B^-\to D^{(*)0}\pi^-$ decays is in general very large and it must be taken into account even at the present level of experimental uncertainty.\\
\begin{acknowledgments}
The author is grateful to Fernando Martinez-Vidal, Abi Soffer and John Walsh for useful comments.
\end{acknowledgments}

\appendix*
\begin{widetext}
\section{Integrals}
\begin{eqnarray}
&&\int_t^{+\infty} |g_\pm(t')|^2 dt' = \frac{1}{2\Gamma}e^{-\Gamma t}\left(\frac{y\sinh(y\Gamma t)+\cosh(y\Gamma t)}{1-y^2}\pm\frac{-x\sin(x\Gamma t)+\cos(x\Gamma t)}{1+x^2}\right)\label{eq:gpgp_gmgm_timecut}\\
&&\int_t^{+\infty} g_+(t')g_-^*(t') dt' = \frac{1}{2\Gamma}e^{-\Gamma t}\left(i\ \frac{\sin(x\Gamma t)+x\cos(x\Gamma t)}{1+x^2}-\frac{\sinh(y\Gamma t)+y\cosh(y\Gamma t)}{1-y^2}\right)\label{eq:gpgm_timecut}
\end{eqnarray}
\end{widetext}


\begin{thebibliography}{00}
\bibitem{babar_gammacomb} P.~del Amo Sanchez {\it et al.} (BaBar Collaboration), Phys. Rev. Lett. {\bf 105}, 121801 (2010).
\bibitem{belle_dalitz_moddep} A.~Poluektov {\it et al.} (Belle Collaboration), Phys. Rev. D {\bf 81}, 112002 (2010). 
\bibitem{lhcb_gammacomb} R.~Aaij {\it et al.} (LHCb Collaboration), Phys. Lett. B{\bf 726}, 151 (2013).
\bibitem{brod_zupan} J.~Brod and J.~Zupan, arXiv:1308.5663.
\bibitem{lhcb_upgrade} R.~Aaij {\it et al.}, Report No. CERN-LHCC-2011-001.
\bibitem{belle2} T.~Aushev {\it et al.}, arXiv:1002.5012.
\bibitem{dmix_gamma0} C.~C.~Meca and J.~P.~Silva, Phys. Rev. Lett. {\bf 81}, 1377 (1998); A.~Amorim, M.~G.~Santos and J.~P.~Silva, Phys. Rev. D {\bf 59}, 056001 (1999); D.~Atwood, I.~Dunietz and A.~Soni, Phys. Rev. D {\bf 63}, 036005 (2001).
\bibitem{dmix_gamma1} J.~P.~Silva and A.~Soffer, Phys. Rev. D {\bf 61}, 112001 (2000).
\bibitem{dmix_gamma2} Y.~Grossman, A.~Soffer and J.~Zupan, Phys. Rev. D {\bf 72}, 031501 (2005).
\bibitem{gamma_cp} W.~Wang, Phys. Rev. Lett. {\bf 110}, 061802 (2013); M.~Martone and J.~Zupan, Phys. Rev. D {\bf 87}, 034005 (2013); B.~Bhattacharya, D.~London, M.~Gronau and J.~L.~Rosner, Phys. Rev. D {\bf 87}, 074002 (2013); A.~Bondar, A.~Dolgov, A.~Poluetkov and V.~Vorobiev, Eur. Phys. J. C {\bf 73}, 2476 (2013).
\bibitem{grossman_savastio} Y.~Grossman and M.~Savastio, arXiv:1311.3575.
\bibitem{hfag12} Y.~Amhis {\it et al.} (Heavy Flavor Averaging Group), arXiv:1207.1158; updated results and plots available at: \url{http://www.slac.stanford.edu/xorg/hfag/}.
\bibitem{lhcb_dalitz} R.~Aaij {\it et al.} (LHCb Collaboration), Phys. Lett. B {\bf 718}, 43 (2012).
\bibitem{ads} D.~Atwood, I~Dunietz and A.~Soni, Phys. Rev. Lett. {\bf 78}, 3257 (1997).
\bibitem{hfag_convention_note} There is a $180^\circ$ difference between the phase $\delta_f$ defined here and the one measured in~\cite{hfag12}.
\bibitem{lhcb_adsglw} R.~Aaij {\it et al.} (LHCb Collaboration), Phys. Lett. B {\bf 712}, 203 (2012); (LHCb Collaboration) Phys. Lett. B {\bf 713}, 351(E) (2012).
\bibitem{ads_kpipi0_babar} J.~P.~Lees {\it et al.} (BaBar Collaboration), Phys. Rev. D {\bf 84}, 012002 (2011).
\bibitem{ads_k3pi_lhcb} R.~Aaij {\it et al.} (LHCb Collaboration), Phys. Lett. B {\bf 723}, 44 (2013).
\bibitem{glw} M.~Gronau and D.~London, Phys. Lett. B {\bf 253}, 483 (1991); M.~Gronau and D.~Wyler, Phys. Lett. B {\bf 265}, 172 (1991).
\bibitem{ggsz} A.~Giri, Y.~Grossman, A.~Soffer and J.~Zupan, Phys. Rev. D {\bf 68}, 054018 (2003). 
\bibitem{bondar_dalitz} A.~Bondar, Proceedings of BINP Special Analysis Meeting on Dalitz Analysis, 24-26 Sep. 2002 (unpublished).
\bibitem{rademacker_wilkinson} J.~Rademacker and G.~Wilkinson, Phys. Lett. B {\bf 647}, 400 (2007).
\bibitem{babar_dalitz_dmix} P.~del Amo Sanchez {\it et al.} (BaBar Collaboration), Phys. Rev. Lett. {\bf 105}, 081803 (2010).
\bibitem{psi3770_conditions} The terms linear in $x$ and $y$ cancel as a consequence of the time-integration and the $D\bar D$ quantum coherence. See for example~\cite{bondar_etal_modind_dmix}.
\bibitem{cleoc_modind} J.~Libby {\it et al.} (CLEO Collaboration), Phys. Rev. D {\bf 82}, 112006 (2010). 
\bibitem{bondar_etal_modind_dmix} A.~Bondar, A.~Poluektov and V.~Vorobiev, Phys. Rev. D {\bf 82}, 034033 (2010).
\bibitem{utfit} UTfit Collaboration, \url{http://www.utfit.org}\,.
\bibitem{ckmfitter} CKMfitter Collaboration, \url{http://ckmfitter.in2p3.fr}\,.
\bibitem{bondar_gershon} A.~Bondar and T.~Gershon, Phys. Rev. D {\bf 70}, 091503 (2004).
\bibitem{gronau_plb557_198} M.~Gronau, Phys. Lett. B {\bf 557}, 198 (2003).
\bibitem{cleoc_deltakpi} D.~M.~Asner {\it et al.} (CLEO Collaboration), Phys. Rev. D {\bf 86}, 112001 (2012).
\end{thebibliography}
\end{document}